\documentstyle[12pt]{article}
\def\lp{\left(}
\def\rp{\right)}
\def\lb{\left[}
\def\rb{\right]}
\def\ai{\'{\i}}

\def\La{\Lambda}

\def\be{\begin{equation}}
\def\ee{\end{equation}}

\begin{document}

\baselineskip.29in

 \title{\Large{\bf Perturbative dynamics of thin-shell wormholes beyond general relativity: an alternative approach}}  

\author{\small Emilio Rub\'{\i}n de Celis$^{1,2,}$\thanks{e-mail: erdec@df.uba.ar}, Cecilia Tomasini$^2$\thanks{e-mail: ctomasini@df.uba.ar}, Claudio Simeone$^{1,2,}$\thanks{e-mail: csimeone@df.uba.ar}\\
{\small $^1$ IFIBA -- CONICET, Ciudad Universitaria Pab. I, 1428,  Buenos Aires, Argentina}\\
{\small $^2$ Departamento de F\'{\i}sica, Facultad de Ciencias Exactas y  Naturales,} \\  
{\small Universidad de Buenos Aires, Ciudad Universitaria Pab. I, 1428,  Buenos Aires, Argentina}}  

\maketitle

\vskip1.5cm

\noindent ABSTRACT: Recent studies relating the approximations for the equations of state for thin shells and their consequent perturbative evolution are extended to thin-shell wormholes in theories beyond general relativity and more than four spacetime dimensions.  The assumption of equations of state of the same form for static and slowly evolving shells appears as a strong restriction excluding the possibility of oscillatory evolutions. Then the new results considerably differ from previous ones obtained within the usual linearized approach. 

\vskip2cm

{\it KEY WORDS:} Dilaton gravity; Einstein--Gauss--Bonnet gravity; thin-shell wormholes.

\vskip1cm

{\it PACS numbers:} 04.40.Nr, 04.50.Kd, 04.50.Gh

\newpage

\section{Introduction}

The outstanding physical properties of traversable wormholes \cite{book} have driven a considerable amount of work. In particular, two central issues were addresed: the kind of matter supporting such topologically non trivial geometries, and their stability under perturbations. A class of wormholes for which these aspects have been studied in great detail are those supported by thin matter layers ({\it thin shells}). Most stability analyses of thin-shell wormholes have been performed within the approach in which a linear relation is assumed between the energy density and pressure; see the leading works \cite{loupo,poivi}, and for recent studies see, for example, \cite{other2,cloud,sh5} and references therein. On the other hand, in a series of articles following Ref. \cite{nos1} the perturbative dynamics of shells in several examples within the framework of relativistic gravity was studied under the assumption that the form of the relations between the energy density and pressures valid for a static shell also holds for the evolution following a slow perturbation; the results obtained within this approach differed significantly from previous ones obtained with the linearized stability treatment (see \cite {ec17} for a detailed discussion).  Different evolutions associated to different equations of state clearly do not imply a flaw in any approach adopted; however, it is natural to compare the results and try to identify an aspect playing a central role in determining  different behaviours. In what follows, we extend the proposal of \cite{nos1} to theories of gravity beyond general relativity and more than four spacetime dimensions. We consider the perturbative dynamics of thin-shells supporting wormholes in the frameworks of four-dimensional dilaton gravity and in five-dimensional Einstein--Gauss--Bonnet gravity. Within the linearized stability analysis all the examples here examined admited static stable configurations \cite{whdil1,whdil2,whgb,hal,whgb4}; in the new approach introduced in \cite{nos1} the results will be notably different. As usual, we choose natural units, so that $c=G_n=1$, with $G_n$ the n-dimensional gravitational constant.

\section{Dilaton wormholes in 3+1 dimensions}

The classical symmetries of the action describing the world-sheet dynamics of strings on a curved manifold in presence of background fields are not preserved at the quantum level in an arbitrary background, unless certain restrictions are imposed on the admissible external fields.  In brief, the requirement of preserving at the quantum level the conformal invariance of the two-dimensional world sheet theory, formulated up to the first order in the inverse of the string tension, leads to the same equations of motion resulting from the variational principle $\delta S=0$ imposed on certain $D$-dimensional field theory.
In particular, in the theory of closed bosonic strings \cite{polch} with the addition of an electromagnetic field, the suitable action $S$ for $D=4$ written in the so-called Einstein frame has the form \cite{gas}
\be  
S=\int d^4x\sqrt{-g}\left(-R+2(\nabla \phi)^2+e^{-2b\phi}F^2\right),\label{emd}
\ee
where $R$ is the Ricci scalar of the background metric $g_{\mu\nu}$, $\phi$ is the dilaton field, $F^{\mu \nu }$ is the electromagnetic field and $b$ $(0\leq b\leq 1)$ determines the coupling between them. The variational principle $\delta S=0$ leads to
\be
\nabla _{\mu }\lp e^{-2b\phi }F^{\mu \nu }\rp =0,
\ee
\be
\nabla ^{2}\phi +\frac{b}{2}e^{-2b\phi }F^{2}=0,
\ee
\be
R_{\mu \nu }=2\nabla _{\mu }\phi \nabla _{\nu }\phi +2e^{-2b\phi }
\lp F_{\mu \alpha }F_{\nu }^{\; \alpha }-\frac{1}{4}g_{\mu \nu }F^{2}\rp .  \label{feq}
\ee
These are the equations for the dilaton and electromagnetic fields, and the Einstein equations with  these  fields as the source. They admit static spherically symmetric solutions with a metric of the form  \cite{GHS,HH,gima}
\be
ds_\pm^2=-f(r)dt^2+f^{-1}(r)dr^2+h(r)(d\theta^2+\sin^2\theta d\varphi^2),
\ee
where
\be
f(r)=\lp 1-\frac{A}{r}\rp\lp 1-\frac{B}{r}\rp^{(1-b^2)/(1+b^2)},
\ee
\be
h(r)= r^2\lp 1-\frac{B}{r}\rp^{2b^2/(1+b^2)}.
\ee
The constants $A$ and $B$ above correspond to the inner and outer horizons of a black hole geometry, and are related with the mass $M$ and electromagnetic charge $Q$ by
\be
A=M+\sqrt{M^2-(1-b^2)Q^2}, \ \ \ \  \ B=\frac{(1+b^2)Q^2}{M+\sqrt{M^2-(1-b^2)Q^2}}
\ee
In what follows we consider a shell placed at a radius $a$ outside the outer horizon of the original black hole manifold. This shell is defined to connect two identical copies of the exterior part of that geometry, so that the complete resulting geometry is that of a wormhole symmetric across the throat, which is a minimal area surface. The matter on the shell is related with the geometry at each side by the Lanczos equations\footnote{We assume the necessary relative factor in the matter contribution to the action to get the usual factor $8 \pi$ on the right-hand side of Eq. (\ref{Lanc}).} \cite{daris,daris1,daris2,daris3}
\be \label{Lanc}
\langle K_{ij}-K h_{ij}\rangle = - 8\pi S_{ij},
\ee
where $S_{ij}$ is the surface energy-momentum tensor, ${K}^{\pm}_{ij}$ is the extrinsic curvature at each side given by
\be
{K}^{\pm}_{ij}=-n^{\pm}_{\alpha}\left(\frac{\partial^{2}X^{\alpha}}{\partial\xi^{i}\partial\xi^{j}}+\Gamma^{\alpha}_{\beta\gamma}\frac{\partial X^{\beta}}{\partial\xi^{i}}\frac{\partial X^{\gamma}}{\partial\xi^{j}}\right)_{r=a},
\ee
$K$ is the corresponding trace and $\langle\cdot\rangle$ stands for the jump of  a given quantity across the surface $r=a$; $n_\alpha^\pm$ are the components of the unit normals at each side. Greek indices label the background coordinates, and latin indices label those on the surface $r=a$.  The resulting energy density and pressure are then  
\be
\sigma=-\frac{1}{4\pi}\frac{h'(a)}{h(a)}\sqrt{f(a)+{\dot a}^2},
\ee
\be
p=\frac{1}{8\pi}\sqrt{f(a)+{\dot a}^2}\lb\frac{2\ddot a+f'(a)}{f(a)+{\dot a}^2}+\frac{h'(a)}{h(a)}\rb,
\ee
where a dot stands for a derivative with respect to the proper time on the shell, and a prime denotes  a derivative with respect to the radial coordinate $r$. A static configuration would then have the energy-momentum given by
\be
\sigma_0=-\frac{\sqrt{f(a_0)}}{4\pi }\frac{h'(a_0)}{h(a_0)},
\ee
\be
p_0=\frac{\sqrt{f(a_0)}}{8\pi }\lb\frac{f'(a_0)}{f(a_0)}+\frac{h'(a_0)}{h(a_0)}\rb.
\ee
Starting from this point, stable static solutions were found for a linearized equation of state \cite{whdil1} and also for a generalized Chaplygin equation of state \cite{whdil2}. Here we will follow the alternative approach introduced in \cite{nos1} and recently analysed in \cite{ec17}. From the equations above we can read the relation existing between the surface energy density and the pressure for the case of a shell evolving with time:
\be
p=Y\sigma, \ \ \ \ \ Y=-\frac{1}{2}\lb 1+\lp\frac{2\ddot a+f'(a)}{f(a)+{\dot a}^2}\rp\frac{h(a)}{h'(a)}\rb,
\ee
and the analogous relation for a static shell:
\be
p_0=X_0\sigma_0, \ \ \ \ \ X_0=-\frac{1}{2}\lb 1+\frac{f'(a_0)h(a_0)}{f(a_0)h'(a_0)}\rb.
\ee
If now we follow the approximate --perturbative-- treatment  introduced in \cite{nos1}, we demand that the form of the equation of state valid for the static configuration holds for the shell undergoing a slow evolution by imposing the condition
\be
Y=X,\ \ \ \  \ \ X=-\frac{1}{2}\lb 1+\frac{f'(a)h(a)}{f(a)h'(a)}\rb.
\ee
It is straightforward to prove that from this condition the equation of motion obtained for the shell radius is
\be
2\ddot a f(a)={\dot a}^2f'(a),
\ee
which has a solution of the  form
\be
\dot a(\tau)=\dot a_0\sqrt{\frac{f(a(\tau))}{f(a_0)}}.
\ee
This kind of solution excludes any oscillatory motion, as the sign of the velocity of the shell is determined by the initial conditions. Then, as obtained in \cite{ec17} and other works following the same approach, only a monotonic motion is possible under the assumptions adopted. In general, because in the parameter range $0\leq b\leq 1$ the metric function $f(r)$ increases with $r$, we would have a decelerated contraction after an inwards perturbation and an accelerated expansion after an outwards perturbation. In the second case the possibility of a shell speed making eventually invalid the perturbative treatment should be adressed. Because $f(r)\to 1$ as $r\to\infty$, the shell speed is bounded, and in an outwards perturbation a small initial speed ensures a subsequent slow evolution only if $f(a_0)$ is not far from unity; this condition is fulfilled if the radius of the initial configuration is considerably larger than the outer horizon radius of the original black hole geometry.
    
\section{Einstein--Gauss--Bonnet wormholes in 4+1 dimensions}

The theory of gravity in five dimensions associated to the Einstein action plus the so-called Gauss--Bonnet terms is the most general metric theory of gravity which leads to  equations of motion of second order \cite{Lovelock}. This theory was extensively studied mainly because it can be obtained within the string theoretical framework \cite{st1,st2,st3,st4}. The Gauss--Bonnet higher order terms in the gravitational action physically correspond to short distance corrections to general relativity. The study of black hole solutions in Einstein--Gauss--Bonnet theory began in the 80's, when the static spherically symmetric solution was found by Boulware and Deser \cite{BD}. After that, Wiltshire obtained the charged black hole geometry in both Maxwell and Born--Infeld electrodynamics \cite{W,W1}.  
The action of the theory, including an electromagnetic field and non vanishing cosmological constant $\La$, reads
\be
S=\int d^{5}x\sqrt{-g}\lb R-2\La-\frac{1}{4}F_{\mu\nu}F^{\mu\nu}+\alpha\lp R_{\alpha\beta\gamma\delta}R^{\alpha\beta\gamma\delta}-4R_{\alpha\beta}R^{\alpha\beta}+R^2\rp\rb.
\ee
where $\alpha$ is a constant of dimensions $(\mathrm{length})^2$. The value of this constant determines the departure of the theory from pure relativity ($\alpha=0$). Because there has been some interest in pure Gauss--Bonnet gravity, and in this case the correct relative sign of the associated terms in the action requires $\alpha>0$, we shall assume this condition.  
The variational principle with the action above gives field equations which include a spherically symmetric static solution of the form 
\be
ds^2=-f(r)dt^2+f^{-1}(r)dr^2+r^2\lp d\theta^2+\sin^2\theta d\chi^2+\sin^2\theta\sin^2\chi d\varphi^2\rp,\label{geomy}
\ee
where the metric function $f(r)$ admits two branches:
\be
f(r)=1+\frac{r^2}{4\alpha}\mp\frac{r^2}{4\alpha}\sqrt{1+\frac{16\alpha M}{\pi r^4}+\frac{8\alpha Q^2}{3r^6}+\frac{4\alpha\Lambda}{3}}.
\ee
For the minus sign we have the normal branch presenting horizons, and thus describing a black hole geometry, while for the plus sign we would have the so-called exotic branch, which includes a naked singularity. However, in the mathematical construction of a wormhole geometry by the usual cut and paste procedure both horizons and singularities are removed  by placing the wormhole throat at a radius greater than the largest horizon radius; hence there would be no reason to avoid the exotic branch within such framework. Moreover, this branch presents the desirable feature of allowing for simple wormhole configurations supported by ordinary  matter (i.e. matter satisfying the energy conditions) even for several parameters set to zero (see below).   

\subsection{Gauss--Bonnet terms as an effective $T_{\mu\nu}$}

The variational principle $\delta S=0$ leads to field equations which can be understood in two ways. In the first one the Gauss--Bonnet contribution is considered as an effective energy-momentum tensor, so that we recover the usual Einstein equations with an additional source:
\be
R_{\mu\nu}-\frac{1}{2}g_{\mu\nu}R+\La g_{\mu\nu}=\frac{1}{2}\lp T_{\mu\nu}^{EM}+T_{\mu\nu}^{GB}\rp,
\ee
where
\be
T_{\mu\nu}^{EM}=  F_{\mu \alpha}F_{\nu }^{\; \alpha}-\frac{1}{4}g_{\mu \nu }F_{\alpha\beta}F^{\alpha\beta},
\ee
and
\begin{eqnarray}
T_{\mu\nu}^{GB}&=&\alpha\left[ 8R_{\alpha\beta}R^{\alpha\ \beta}_{\ \mu\ \nu}-4R_{\mu\alpha\beta\gamma}R_\nu^{\ \alpha\beta\gamma}\right.\nonumber\\
& &\left. +\ 8R_{\mu\alpha}R^\alpha_{\ \nu}-4RR_{\mu\nu}+g_{\mu\nu} \lp R_{\alpha\beta\gamma\delta}R^{\alpha\beta\gamma\delta}
-4R_{\alpha\beta}R^{\alpha\beta}+R^2\rp\right].
\end{eqnarray}
In this picture the junction conditions at a given surface are just those of general relativity,  
\be
\langle K_{ij}-K h_{ij}\rangle = - 8 \pi S_{ij},
\ee
and the surface energy-momentum tensor includes a Gauss-Bonnet contribution. If the surface joins two identical copies of the exterior part of a spherically symmetric static metric, the resulting new complete manifold is that of a wormhole of throat radius $a$, supported by a shell with energy density and pressure given by
\be
\sigma=-\frac{3}{4\pi a}\sqrt{f(a)+{\dot a}^2},
\ee
\be
p=-\frac{2}{3}\sigma+\frac{1}{8\pi}\frac{2\ddot a+f'(a)}{\sqrt{f(a)+{\dot a}^2}}.
\ee
We see that there is no possibility to avoid exotic matter in the construction of a thin shell wormhole with this interpretation of the Gauss-Bonnet terms. For a static configuration $(a=a_0)$ we have
\be
\sigma_0=-\frac{3\sqrt{f(a_0)}}{4\pi a_0},
\ee
\be
p_0=-\frac{2}{3}\sigma_0+\frac{1}{8\pi}\frac{f'(a_0)}{\sqrt{f(a_0)}}.
\ee
Starting from this point, the existence of static configurations stable under perturbations preserving the symmetry were found in \cite{whgb} following the linearized approch. Here, instead, we follow the procedure of the preceding section. The relation between the static energy density and pressure is  
\be
p_0=X_0\sigma_0, \ \ \ \ \ X_0= -\frac{1}{3}\left(2+\frac{a_0f'(a_0)}{2f(a_0)}\right).
\ee
On the other hand, for the general case of a moving shell at the wormhole throat we have  
\be
p=Y\sigma, \ \ \ \ \ Y=  -\frac{1}{3}\left(2+\frac{2a\ddot a +a f'(a)}{2f(a)+2{\dot a}^2}\right).\label{32}
\ee
If we demand that the form of the equation of state for the static case is preserved in a slow evolution of the shell, we impose the approximation
\be
p=X\sigma, \ \ \ \ \ X= -\frac{1}{3}\left(2+\frac{af'(a)}{2f(a)}\right).\label{33}
\ee
From (\ref{32}) and (\ref{33}) we immediately obtain the condition
\be
f'(a){\dot a}^2=2f(a)\ddot a
\ee
which is the equation of motion for the shell at the wormhole throat. As before, the solution turns to be of the form
\be
\dot a (\tau)=\dot a_0 \sqrt{\frac{f(a(\tau))}{f(a_0)}},
\ee
which implies, again, a monotonic evolution: an expansion if the initial velocity points outwards, and a contraction if it points inwards. The accelerated or decelerated character of the motion would be determined, in each case, by the dependence of the metric function $f$ with the radial coordinate. For example, consider the assumption $\alpha\ll a^2$, so that the theory implies a slight departure from relativity at the scale of our wormhole configuration, and  the simplest case of vanishing charge and vanishing cosmological constant. Then we would have $f(r)\simeq 1-2M/(\pi r^2)$  for the normal branch of the metric, and $f(r)\simeq 1+2M/(\pi r^2)+r^2/(2\alpha)$ for the exotic branch. In the first case the shell would undergo an accelerated expansion with decreasing acceleration after an outwards perturbation, and a decelerated contraction after an inwards perturbation; the same considerations of the preceding examples would then apply regarding the validity of the perturbative approximation. In the second case the assumption $\alpha\ll a^2$ would fail after an inwards perturbation, as there would be no limit for the decrease of the shell radius. On the other hand, for the exotic branch of the metric the theory in this approximation provides a sort of effective cosmological constant $\Lambda_{\mathrm Eff}\sim \alpha^{-1}$ which would drive an outwards expansion with unbounded speed; in fact, for large values of the shell radius we would have $\dot a\sim a$.  This would make eventually invalid the perturbative treatment, which relies in the hypothesis of a slow motion of the shell. 

\subsection{Gauss--Bonnet terms as a geometric object}

 The Gauss--Bonnet terms of the equations of the theory can be associated to the geometry, and not to an effective energy momentum tensor, which allows for a more satisfactory understanding of the character of the matter on the shell joining two metrics. Besides, this has other advantages in the framework in which we are working. In this picture the junction conditions are generalized to include the Gauss--Bonnet contribution, and this has two positive consequences:  First, for certain values of the parameters, thin-shell wormholes can be supported by matter satisfying the energy conditions \cite{whgb2,maeno,hal,whgb3}; moreover, when associated to the exotic branch of Wiltshire solution, wormholes are supported by normal matter  even in the case of a relatively small and positive Gauss--Bonnet constant, and with vanishing charge and zero cosmological constant \cite{whgb3}. Second, as this substantially changes the form of the surface energy density and pressure on the shell, it could allow for an, in principle, different kind of motion. The field equations read
\be
R_{\mu\nu}-\frac{1}{2}g_{\mu\nu}R+ \Lambda g_{\mu\nu}+2\alpha H_{\mu\nu}= 8 \pi \,T_{\mu\nu},\label{field}
\ee
where
\begin{eqnarray}
H_{\mu\nu} & = & RR_{\mu\nu}-2R_{\mu\alpha}R^{\alpha}_{\nu}-2R^{\alpha\beta}R_{\mu \alpha\nu\beta}\nonumber\\
& & +\ R_{\mu}^{\alpha\beta\gamma}R_{\nu\alpha\beta\gamma}-\frac{1}{4}g_{\mu\nu}(R^{2}-4R^{\alpha\beta}R_{\alpha\beta}+R^{\alpha\beta\gamma\delta}R_{\alpha\beta\gamma\delta}).
\end{eqnarray}
The matching conditions relating the metrics at the two sides of a given surface  with the character of matter on this surface  are given by the  Darmois--Israel conditions generalized  to  Einstein--Gauss--Bonnet gravity. They were obtained in Ref. \cite{4} (see also \cite{mae})  starting from the equations above and read
\be
\langle K_{ij}-K h_{ij}\rangle + 2\alpha \langle 3J_{ij}-Jh_{ij}+2P_{iklj}K^{kl}\rangle =- 8 \pi S_{ij},
\ee
where latin indices label the coordinates on the joining surface, and the divergence-free part of the Riemann tensor $P_{ijkl}$ and the tensor $J_{ij}$ are defined as follows:
\be
P_{ijkl}=R_{ijkl}+(R_{jk}h_{li}- R_{jl}h_{ki})-(R_{ik}h_{lj}- R_{il}h_{kj})+\frac{1}{2}R(h_{ik}h_{lj}-h_{il}h_{kj}),
\ee
\be
J_{ij}=\frac{1}{3}\left[2KK_{ik}K^{k}_{j}+K_{kl}K^{kl}K_{ij}-2K_{ik}K^{kl}K_{lj}-K^{2}K_{ij}\right].
\ee
Working with this interpretation of the Einstein--Gauss--Bonnet theory and within the linearized stability approach, the possibility of stable thin-shell wormholes connecting two geometries of the form (\ref{geomy}) was shown in Refs. \cite{hal,whgb4} for suitable values of the parameters. We follow, instead, the procedure of the preceding sections to perform the perturbative dynamics analysis under the approximation introduced in \cite{nos1}. From the junction conditions above, the energy density and pressure for a shell connecting two identical spherically symmetric exterior geometries are  
\be
\sigma=-\frac{1}{8\pi}\lb\frac{6\Delta}{a}-\frac{2\alpha}{a^3}\lp 4\Delta^3-12\Delta(1+{\dot a}^2)\rp\rb,
\ee
\be
p=\frac{1}{8\pi}\lb\frac{4\Delta}{a}+2\frac{\ell}{\Delta}-\frac{8\alpha}{a^2}\lp\ell\Delta-\frac{\ell}{\Delta}(1+{\dot a}^2)-2\Delta\ddot a\rp\rb,
\ee
where we have introduced  
\be
\Delta=\sqrt{f(a)+{\dot a}^2},\ \ \ \ \  \ \ell=\frac{f'(a)}{2}+\ddot a.
\ee
The corresponding expressions for the static case are
\be
\sigma_0=-\frac{\sqrt{f(a_0)}}{8\pi a_0}\lb 6-\frac{4\alpha}{a_0^2}\lp 2f(a_0)-6\rp\rb,
\ee
\be
p_0=\frac{\sqrt{f(a_0)}}{8\pi a_0}\lb 4+\frac{a_0f'(a_0)}{f(a_0)}-4\alpha \frac{f'(a_0)}{a_0}\lp\frac{f(a_0)-1}{f(a_0)}\rp\rb,
\ee
so that the following relation between the energy density and pressure holds:
\be
p_0=X_0\sigma_0,\ \ \  \ \ X_0=\frac{a_0\lb 4a_0f(a_0)+a_0^2f'(a_0)-4\alpha f'(a_0)\lp f(a_0)-1\rp\rb}{f(a_0)\lb8\alpha\lp f(a_0)-3\rp-6a_0^2\rb}.
\ee
In the case of a moving shell, instead, the corresponding relation is
\be
p=Y\sigma,\ \ \ \  \ \ \ Y=\frac{2\Delta a+\ell a^2\Delta^{-1}-4\alpha\lb\ell\Delta-\ell \Delta^{-1}(1+{\dot a}^2)-2\Delta\ddot a\rb}{-3\Delta a+\alpha a^{-1}\lp4\Delta^3-12(1+{\dot a}^2)\Delta\rp}.
\ee
The assumption that the form of the equation of state for a static configuration is a good approximation for a moving shell imposes the condition
\be
Y=X,\ \ \ \ \ \ X=\frac{a\lb 4af(a)+a^2f'(a)-4\alpha f'(a)\lp f(a)-1\rp\rb}{f(a)\lb8\alpha\lp f(a)-3\rp-6a^2\rb}.
\ee
Now, differing from the cases considered above, this condition does not lead in a straightforward way to a simple equation of motion for the shell. However, if we are interested in the possibility of oscillatory evolutions which can be associated to a kind of stable equilibrium, we can look for the existence of turning ``points'', that is values of the shell radius for which it must be $\dot a=0$ and  $\ddot a >0$, or $\dot a=0$ and $\ddot a<0$. At such positions of the shell we would have, under the approximation adopted, that 
\be
X(a)=Y(a,\dot a=0,\ddot a).
\ee
A direct calculation then leads to the condition
\be
\ddot a\lb a^2+4\alpha\lp 1+f(a)\rp\rb=0.
\ee
In a construction starting from the exotic branch of the Wiltshire solution we have always $f(a)>0$, and in the case of the black hole geometry we only consider a shell radius larger than which would be the outer horizon radius, so that again  $f(a)>0$; hence for $\alpha >0$ this condition can be fulfilled only if
\be
\ddot a=0,   
\ee
which contradicts the existence of a turning point. Therefore, despite the considerably greater complexity of the equations of motion in the picture in which the Gauss--Bonnet contribution is associated to the geometry, the outcome is analogous to the conclusion of the simpler approach of the preceding section, and oscillatory evolutions can not take place under the hypothesis adopted.

\section{Summary}

In the present work we have extended the approach to treat the perturbative dynamics of shells first adopted  in Refs. \cite{nos1}, and recently analysed in detail in \cite{ec17}, to different thin-shell wormholes in theories of gravity beyond general relativity. We have considered the shells supporting four-dimensional wormholes in the framework of dilaton gravity and five-dimensional wormholes in two different pictures of Einstein--Gauss--Bonnet gravity.  In all cases, and differing from previous results obtained within the linearized stability procedure \cite{whdil1,whdil2,whgb,hal,whgb4},  the same kind of evolution of  the examples in \cite{nos1,ec17} and related works has been found. While in some cases there is at least the possibility of slow evolutions, in other cases the shells at the wormhole throats would even speed up in a way making eventually not valid the perturbative treatment.  The approximation of relations between the energy density and pressure of the same form for static shells and for shells undergoing a slow symmetric perturbation then appears as a strong restriction excluding the possibility of oscillatory solutions associated to stable wormhole configurations.  

\section*{Acknowledgments}

This work has been supported by CONICET and Universidad de Buenos Aires.

\end{document}